\documentclass[10pt,twocolumn]{article}

\usepackage{graphicx}

\begin{document}

\normalsize

\parbox{16.5cm}{
\begin{center}
{\large \bf Electronic and magnetic properties of GaMnAs: Annealing effects}

\vspace{0.5cm}
\underbar{W. Limmer}$^{1\ast}$, A. Koeder$^{1}$, S. Frank$^{1}$, M. Glunk$^{1}$, W. Schoch$^{1}$,
V. Avrutin$^{1}$, K. Zuern$^{2}$, \\ R. Sauer$^{1}$, and A. Waag$^{1}$\\
$^{1}$Abteilung Halbleiterphysik, Universit\"at Ulm, D-89069 Ulm, Germany\\
$^{2}$Abteilung Festk\"orperphysik, Universit\"at Ulm, D-89069 Ulm, Germany\\
$^{\ast}$Corresponding author: Phone:+49-731-50-26130 \hspace{0.4cm}
Fax:+49-731-50-26108 \hspace{0.4cm} Email:wolfgang.limmer@physik.uni-ulm.de
\end{center}

\hrulefill

\vspace{0.5cm}
{\bf Abstract}:
The effect of short-time and long-time annealing at 250$^\circ$C on the conductivity, hole density, and Curie
temperature of GaMnAs single layers and GaMnAs/InGaMnAs heterostructures is studied by in-situ conductivity
measurements as well as Raman and SQUID measurements before and after annealing. Whereas the conductivity monotonously
increases with increasing annealing time, the hole density and the Curie temperature show a saturation after
annealing for 30 minutes. The incorporation of thin InGaMnAs layers drastically enhances the Curie temperature
of the GaMnAs layers.

\vspace{0.5cm}
{\it PACS}: 75.50.Pp, 72.80.-r, 78.30.Fs\\
{\it Keywords}: GaMnAs, annealing, conductivity, Curie temperature, carrier density

\hrulefill }

\vspace{0.3cm}
{\bf 1. Introduction}

\vspace{0.2cm}
\hspace*{0.3cm}
The III-V semimagnetic semiconductor GaMnAs is currently under intense
investigation in the field of spin electronics. Its magnetic properties arise from
the magnetic moments of the Mn atoms which act as acceptors on the Ga lattice site.
The ferromagnetism is mediated by delocalized or weakly localized holes and the Curie
temperature $T_C$ is suggested to depend on both, the Mn content $x$ and the hole
concentration $p$ according to [1]:
\begin{equation}
T_C\propto x \times p^{1/3}\;.
\end{equation}
It has been shown that post-growth annealing of GaMnAs at sufficiently low temperatures
leads to an increase of $T_C$, depending on the annealing temperature and on the annealing time [2-5].
Values of $T_C$ as high as 150 K have been reported so far [2], and the enhancement of $T_C$
has been usually traced back to a removal of compensating defects, and thus, to
an increase of the hole concentration following the relation given in Eq.~(1).
Potashnik et al. have shown that annealing at 250$^\circ$C for less than 2 h significantly enhances
the conductivity and $T_C$, but continuing the annealing for longer times suppresses both [3].
Edmonds et al. have annealed GaMnAs thin films at 175$^\circ$C for more than 100 hours,
monitoring the resistivity during the annealing, and found a monotonous increase of the conductivity with
increasing annealing time as well as a linear relation between $T_C$ and the room-temperature (RT)
electrical conductivity [4]. These results suggest the presence of at least two competing thermally
activated processes, leading, respectively, to an increase and a decrease of the conductivity and $T_C$.
Moreover, the RT conductivity seems to guide the way to an optimization of $T_C$.
\newline

\vspace*{9.0cm}
In this work, we investigate the effect of short-time and long-time annealing at 250$^\circ$C
on the electronic and magnetic properties of GaMnAs by in-situ measurements of the electrical
conductivity, by micro-Raman spectroscopy, and by superconducting quantum interference device
(SQUID) magnetization measurements. Furthermore, we demonstrate that the incorporation of thin
InGaMnAs films into the GaMn\-As material leads to a drastic enhancement of the Curie temperature.

\vspace{0.3cm}
{\bf 2. Experimental details}

\vspace{0.2cm}
\hspace*{0.3cm}
GaMnAs epilayers and GaMnAs/InGaMnAs heterostructures were grown by low-temperature
molecular-beam epitaxy (MBE) on semi-insulating GaAs(001) substrates at 230$^\circ$C using solid
metal sources and a III/V beam equivalent pressure ratio of 1/10. For the heterostructures,
sequences of sixteen 20-nm-thick GaMnAs barrier layers separated by 6-nm-thick InGaMnAs quantum films
were grown on high-temperature GaAs buffer layers. The Mn content in the GaMnAs single layers and in the
GaMn\-As barriers was determined from high-resolution x-ray diffraction (HRXRD) and
amounted to 5.8~$\%$. The In content in the InGaMnAs films was varied from sample to sample between 15
and 40~$\%$. The HRXRD measurements revealed sharp GaMn\-As/InGaMnAs interfaces.
For the electrical measurements, Hall bars with conventional AuGe contacts were prepared on several
pieces of the cleaved samples. The contacts were checked to be ohmic with negligibly low resistance.
The samples were annealed in air using a LINKAM THMS 600 heating chamber
equipped with an electrical feed through, which enabled us to perform in-situ measurements of the conductivity.
The samples were mounted on a silver block which could be heated electrically or cooled by liquid nitrogen
over the temperature range from \linebreak -200$^\circ$C to 300$^\circ$C within 2 minutes.
Micro-Raman measurements were performed at RT using the 514-nm line of
an Ar$^{+}$ laser as an excitation source. The corresponding light-penetration depth can be estimated by
$1/\alpha \approx 100$ nm, where $\alpha$ denotes the optical absorption coefficient. The laser
beam was focused onto the sample surface by a microscope lens system yielding a spot diameter of 0.7 $\mu$m. The
Raman signals were detected in the backscattering configuration $\bar z(x,y)z$
using a DILOR XY 800 mm triple-grating spectrometer with a confocal
entrance optics and a LN$_{2}$-cooled charge-coupled device detector.
The magnetization measurements were carried out in a QUANTUM DESIGN MPMS 5 SQUID magnetometer
applying an in-plane magnetic field of 5 mT.

\vspace{0.3cm}
{\bf 3. Results and discussion}

\vspace{0.2cm}
\hspace*{0.3cm}
Most of the experimental data presented in this work were recorded from a 1-$\mu$m-thick GaMn\-As
epilayer. Qualitatively similar results were obtained for a 300-nm-thick GaMnAs epilayer, grown at
200$^\circ$C, and for the GaMnAs/InGaMnAs heterostrucures.

In order to study the effect of short-time annealing on the electrical conductivity, we have annealed
a 1-$\mu$m-thick GaMnAs epilayer at successively elevated temperatures. The annealing temperature was raised from
50$^\circ$C to 300$^\circ$C in steps of 25$^\circ$C, maintaining an annealing time of 5 minutes for each
temperature. The conductivity at RT was probed by rapidly lowering the temperature to 25$^\circ$C between
successive annealing steps. Figure~1 depicts the in-situ monitored conductivity as a function of time.
\begin{center}
\includegraphics[width=8.0cm,height=5.7cm]{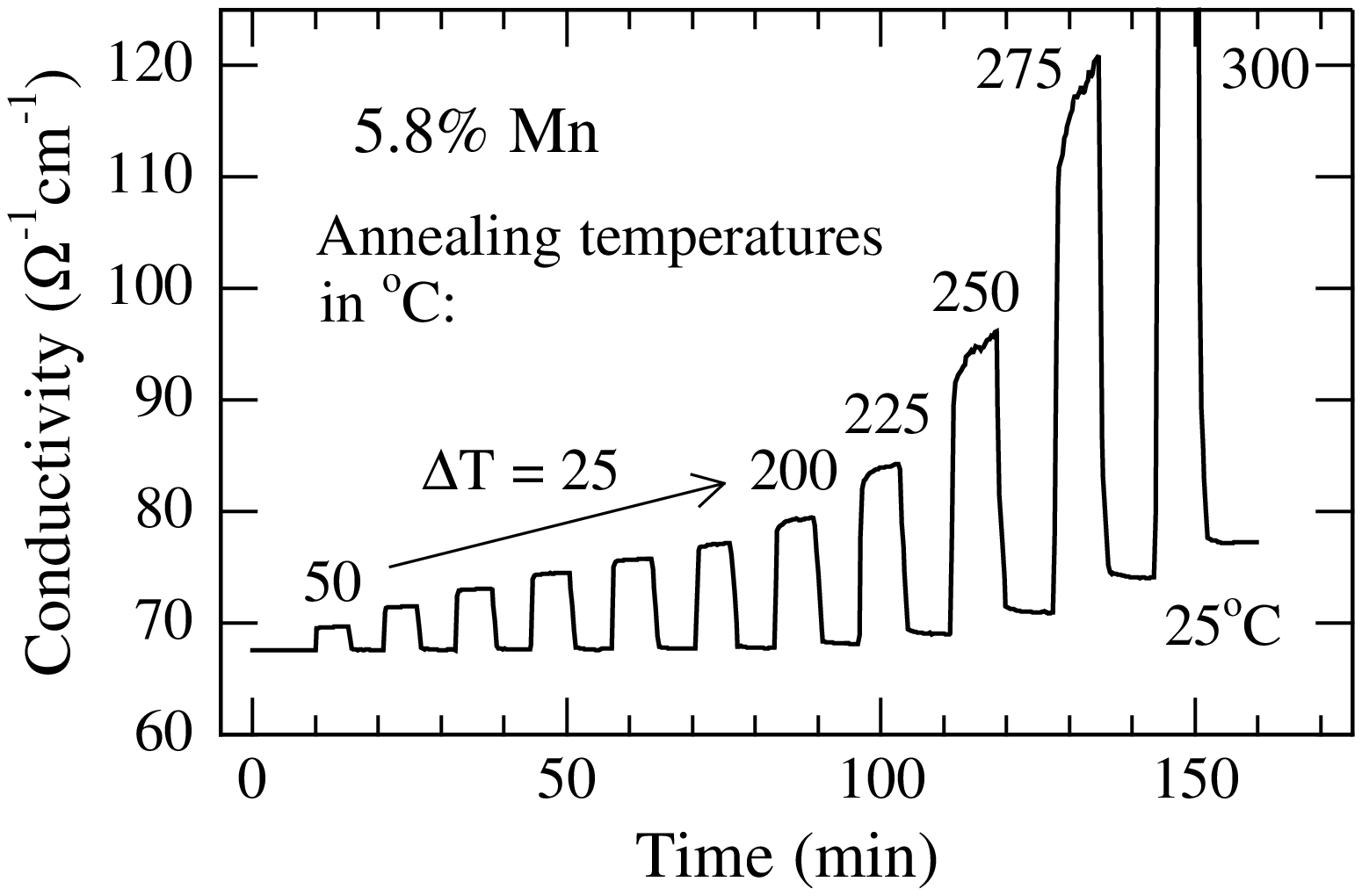}
{\bf Fig.~1}: In-situ monitoring of the conductivity during short-time annealing at successively
elevated annealing temperatures.
\end{center}
No significant change in the RT conductivity is observed for annealing temperatures below 200$^\circ$C.
It should be mentioned that this does not hold for annealing times much longer than that used in our
experiment [4]. Annealing at 200$^\circ$C or at higher temperatures results in a pronounced increase of
the RT conductivity, indicating a rise in the hole concentration or mobility or both.

The effect of long-time annealing at 250$^\circ$C on the conductivity is shown in Fig.~2,
where the evolution of the conductivity is monitored by an in-situ measurement for 400 minutes.
The conductivity monotonously rises with a rate that gradually decreases with increasing annealing time.
After total annealing times of 30, 60, 180, and 370 minutes, the annealing process was interrupted for
10 minutes and the temperature was rapidly lowered to 25$^\circ$C in order to probe the RT conductivity
(dashed line). Its dependence on annealing time is similar to that of the conductivity at 250$^\circ$C.

\vspace{-0.3cm}
\begin{center}
\includegraphics[width=8.0cm,height=5.7cm]{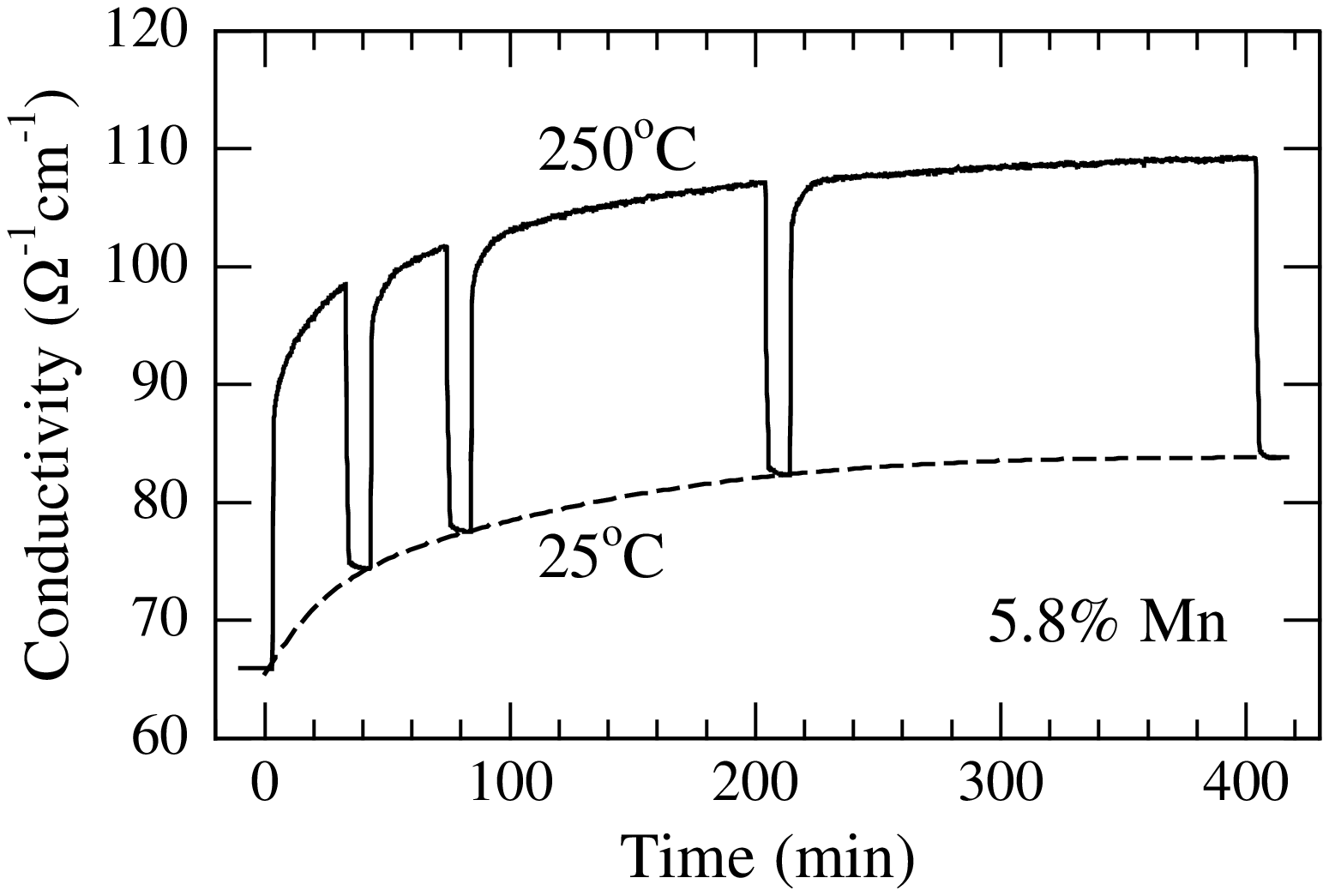}
{\bf Fig.~2}: In-situ monitoring of the conductivity during long-time annealing at 250$^\circ$C.
\end{center}
During the first 30 minutes the RT conductivity increases from 66 to 74 $\Omega^{-1}$cm$^{-1}$.
After annealing for 370 minutes the RT conductivity has almost saturated at a value of 84 $\Omega^{-1}$cm$^{-1}$.
In contrast to Ref.~[3], no decrease of the conductivity for annealing times longer than 2 h is observed.
If we assume the RT hole mobility to remain constant throughout the annealing process, an increase of the
hole concentration by a factor of 1.12 during the first 30 minutes, and by a factor of 1.27 during the whole
process is deduced.

\vspace{-0.9cm}
\begin{center}
\includegraphics[width=8.5cm,height=6.5cm]{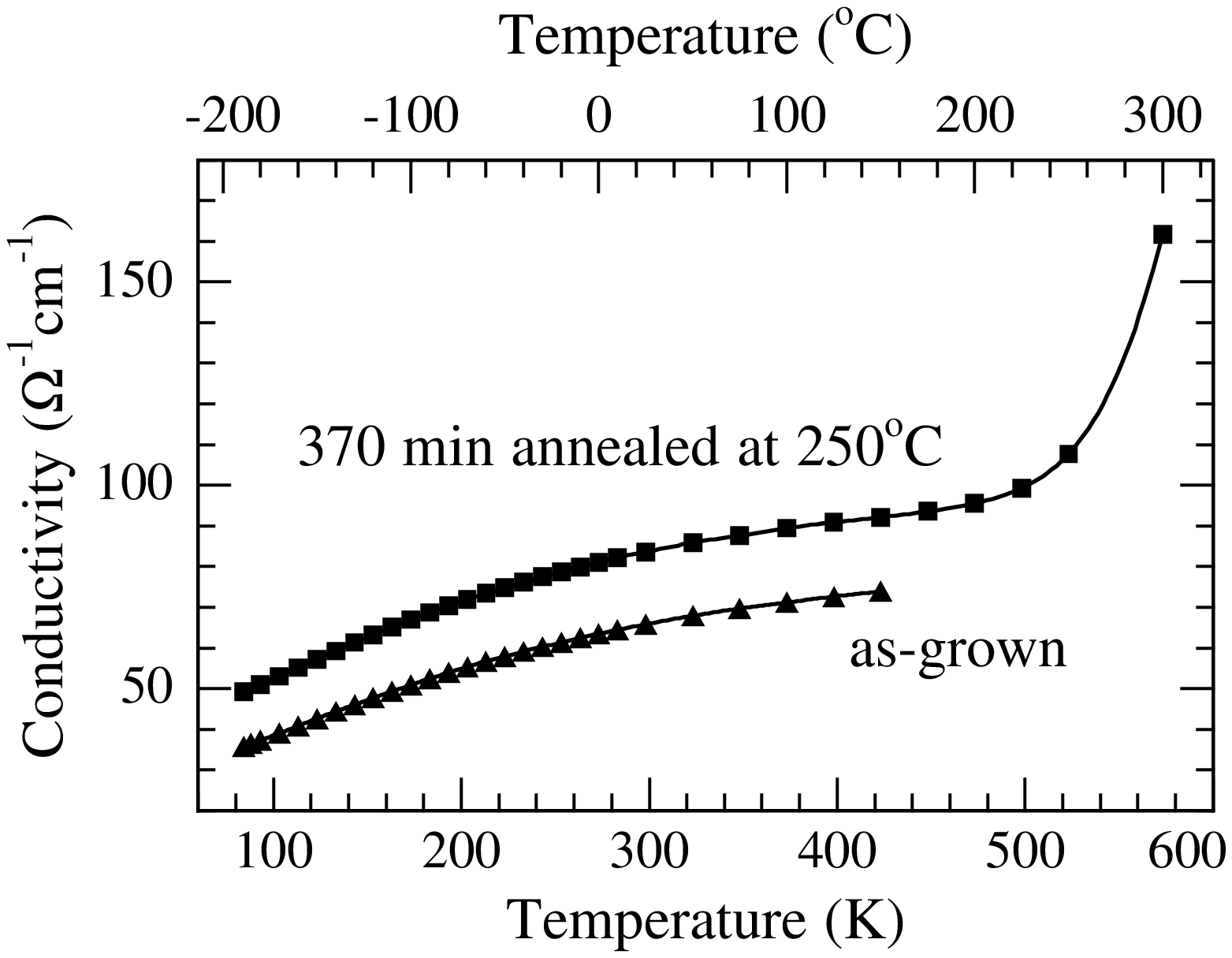}
{\bf Fig.~3}: Temperature-dependent conductivity of a GaMnAs layer before and after annealing.
\end{center}
The conductivity as a function of temperature is shown in Fig.~3 for the as-grown and for the
370-min-annealed sample in the temperature range from -200$^\circ$C to 150$^\circ$C and from
-200$^\circ$C to 300$^\circ$C, respectively. Below 200$^\circ$C the conductivity shows the typical behavior
for metallic GaMnAs, namely a monotonous increase with rising temperature. However, above 200$^\circ$C
the onset of an unexpected exponential increase is observed. Recently, a similar effect has been reported
in Ref.~[6], but at present the origin of this drastic increase is not yet clear.
An annealing effect by raising the temperature of the annealed sample to 300$^\circ$C for only 3 minutes
can be excluded since the RT value of the conductivity has remained unchanged.

Hole concentrations in GaMnAs can be estimated from Raman scattering by coupled plasmon-LO-phonon modes [7].
In Fig.~4 Raman spectra, recorded from the sample examined above, are shown before and after annealing at
250$^\circ$C for 30 and 370 minutes. As discussed in detail in Ref.~[7], the high hole concentration in
GaMnAs leads to the formation of a phonon-like coupled mode of the longitudinal optical (LO) phonon and
the overdamped hole plasmon. With increasing hole concentration this mode shifts from the frequency of the
LO phonon to that of the transverse optical (TO) phonon. Thus, from a comparison of the Raman spectra shown in
Fig.~4, a significant increase of the hole concentration after annealing can be deduced.
\begin{center}
\includegraphics[width=8.0cm,height=6.8cm]{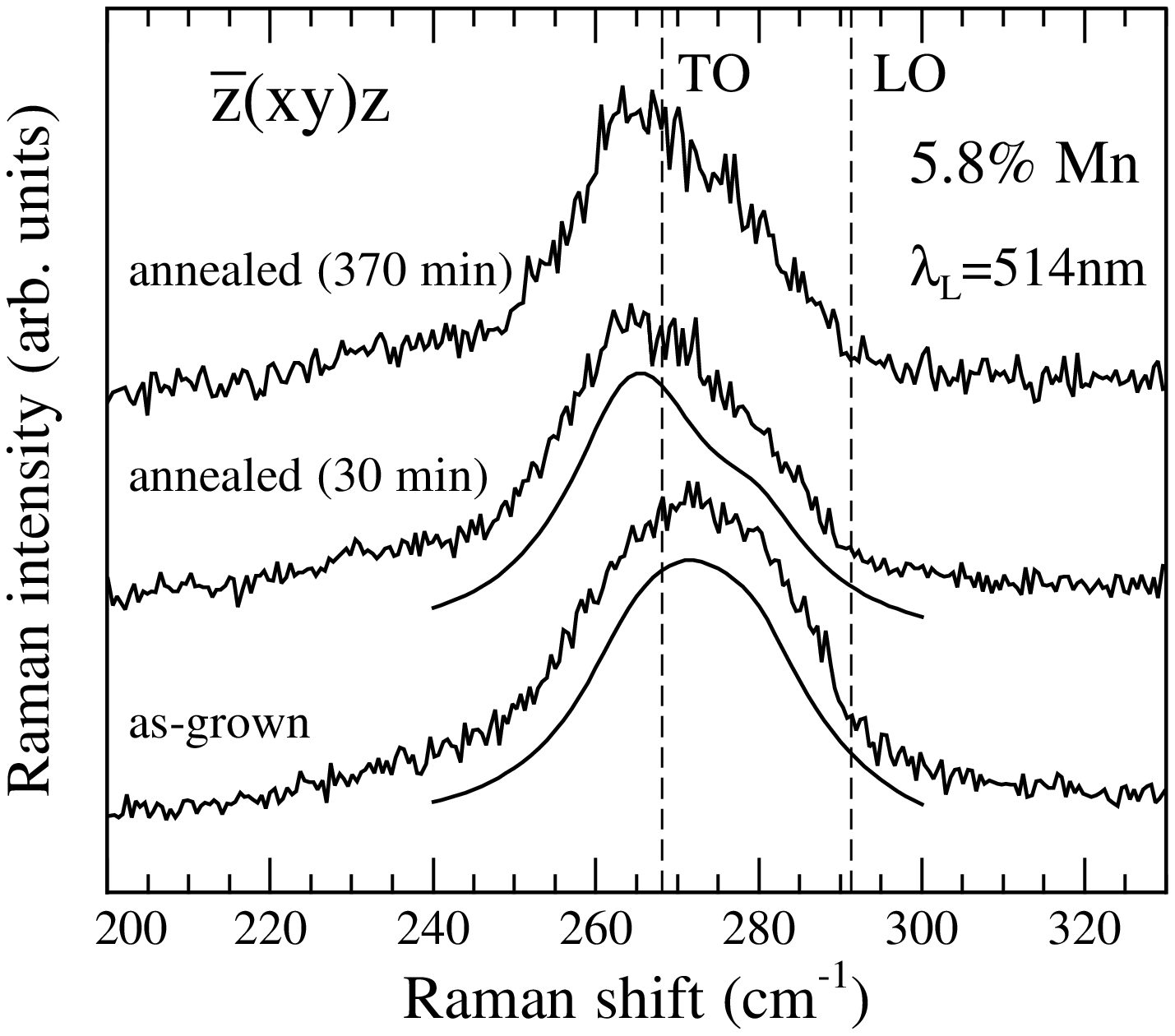}
{\bf Fig.~4}: Raman spectra recorded from a GaMnAs layer before and after
annealing at 250$^\circ$C. The solid lines are calculated line shapes.
\end{center}
The Raman spectrum recorded from the 370-min-annealed sample does not significantly
differ from that of the 30-min-annealed sample. Therefore, according to the Raman spectra,
no further increase of the hole concentration occurs for annealing times longer than 30 minutes,
in contrast to our conductivity measurements, but in aggreement with the results reported in Ref.~[3].
The solid lines, offset for clarity, represent model calculations of the Raman line shapes yielding the hole
concentrations $3.5\times 10^{20}$ cm$^{-3}$ and $11.0\times 10^{20}$ cm$^{-3}$ for the
as-grown and the annealed samples, respectively [7].
Quantitatively, the increase of the hole concentration within the first 30 minutes of annealing
by a factor of 3.14 disagrees with the increase of the RT conductivity, where a factor of 1.12 has been determined
by the electrical measurements.
We explain the discrepancy between the results obtained from the Raman spectra and the conductivity measurements
by vertical gradients of the hole concentration and the hole mobility in the GaMnAs layer.
Electrochemical capacitance-voltage measurements have shown that near the surface the
hole concentration is much higher than in the bulk [8]. Therefore, the Raman signal which primarily stems from the
surface region yields a higher hole density than electrical measurements which probe the whole GaMnAs layer.
Moreover, we suggest the hole densities and hole mobilities near the surface and in the bulk to be affected
by the annealing process in a different way.

According to Eq.~(1), an increase of the hole concentration should result in an increase of the Curie temperature.
Figure~5 depicts SQUID magnetization measurements recorded from the as-grown and from two annealed GaMnAs layers.
\begin{center}
\includegraphics[width=8.0cm,height=5.7cm]{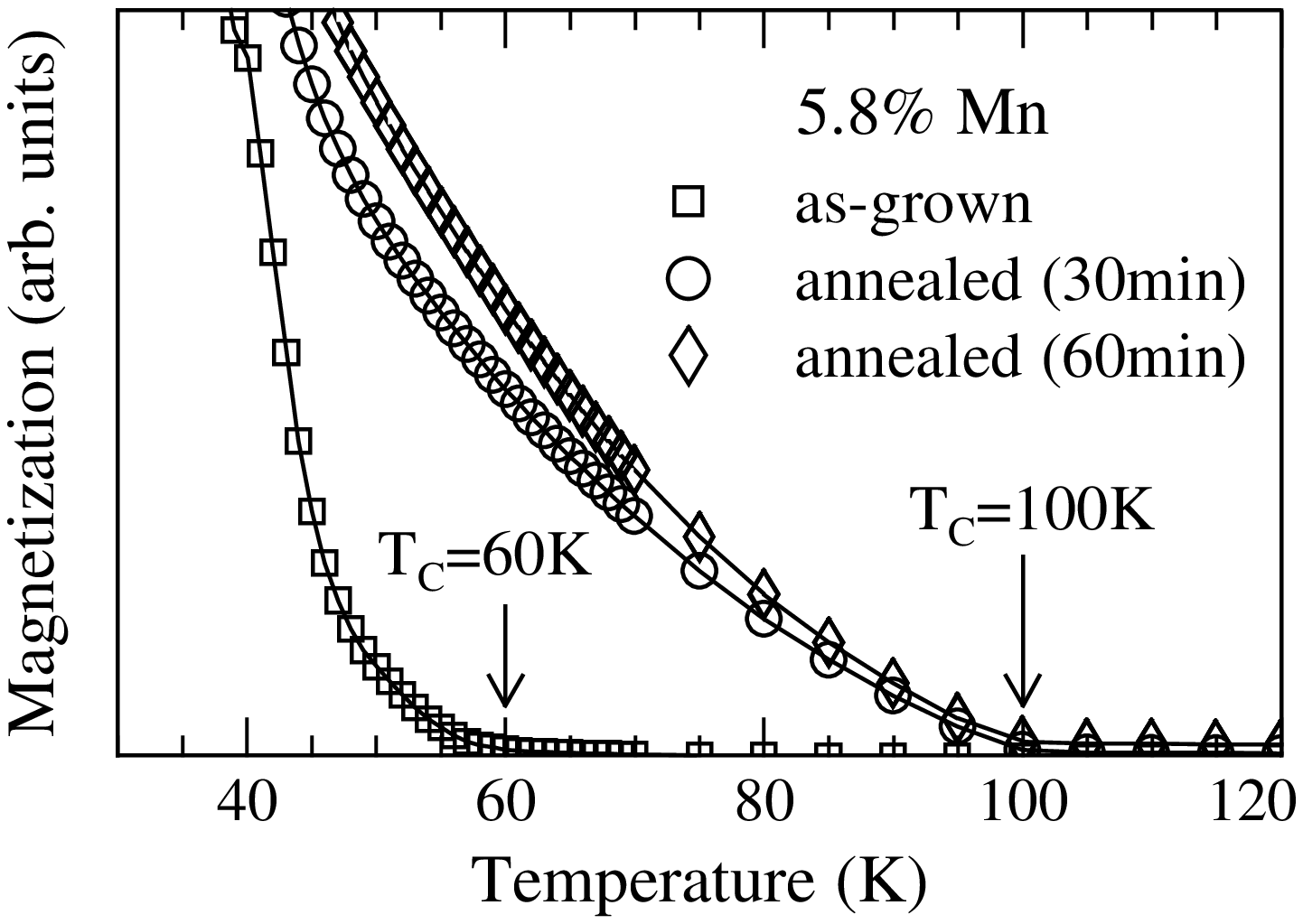}
{\bf Fig.~5}: Magnetization of a GaMnAs layer as a function of temperature before and after annealing.
\end{center}
A Curie temperature $T_C$ of 60 K is deduced for the as-grown sample and a constant value of 100 K is obtained for
the two samples annealed at 250$^\circ$C for 30 and 60 minutes, respectively. This result is in good agreement with that
obtained from the Raman measurements. No further enhancement of the hole concentration or Curie temperature for
annealing longer than 30 minutes is observed by both methods. Moreover, according to Eq.~(1) an increase of the
hole concentration by a factor
of 3.14, as deduced from the Raman spectra, should result in an enhancement of $T_C$ by a factor of 1.46. Actually, a
factor of 1.67 is determined from the SQUID measurements. The small discrepancy between the two values may be explained
by a reduction of the number of antiferromagnetically ordered Mn atoms during annealing
[9]. We suggest that the small slopes of the magnetization curves below $T_C$ in Fig.~5 result from
the vertical gradient of the hole density in the GaMnAs layer. Therefore, the Curie
temperatures indicated in Fig.~5 have to be attributed to the near-surface region, in
accordance with the hole densities obtained from the Raman measurements.

The incorporation of thin InGaMnAs films into the GaMnAs layer leads to a pronounced increase of the
Curie temperature. This is demonstrated in Fig.~6, where the magnetization curves of a GaMn\-As/InGaMnAs
heterostructure with 40 $\%$ In are shown before and after annealing at 250$^\circ$C for 30 minutes.
The curve of the as-grown GaMnAs single layer is drawn by a dashed line for comparison.
\begin{center}
\includegraphics[width=8.0cm,height=5.7cm]{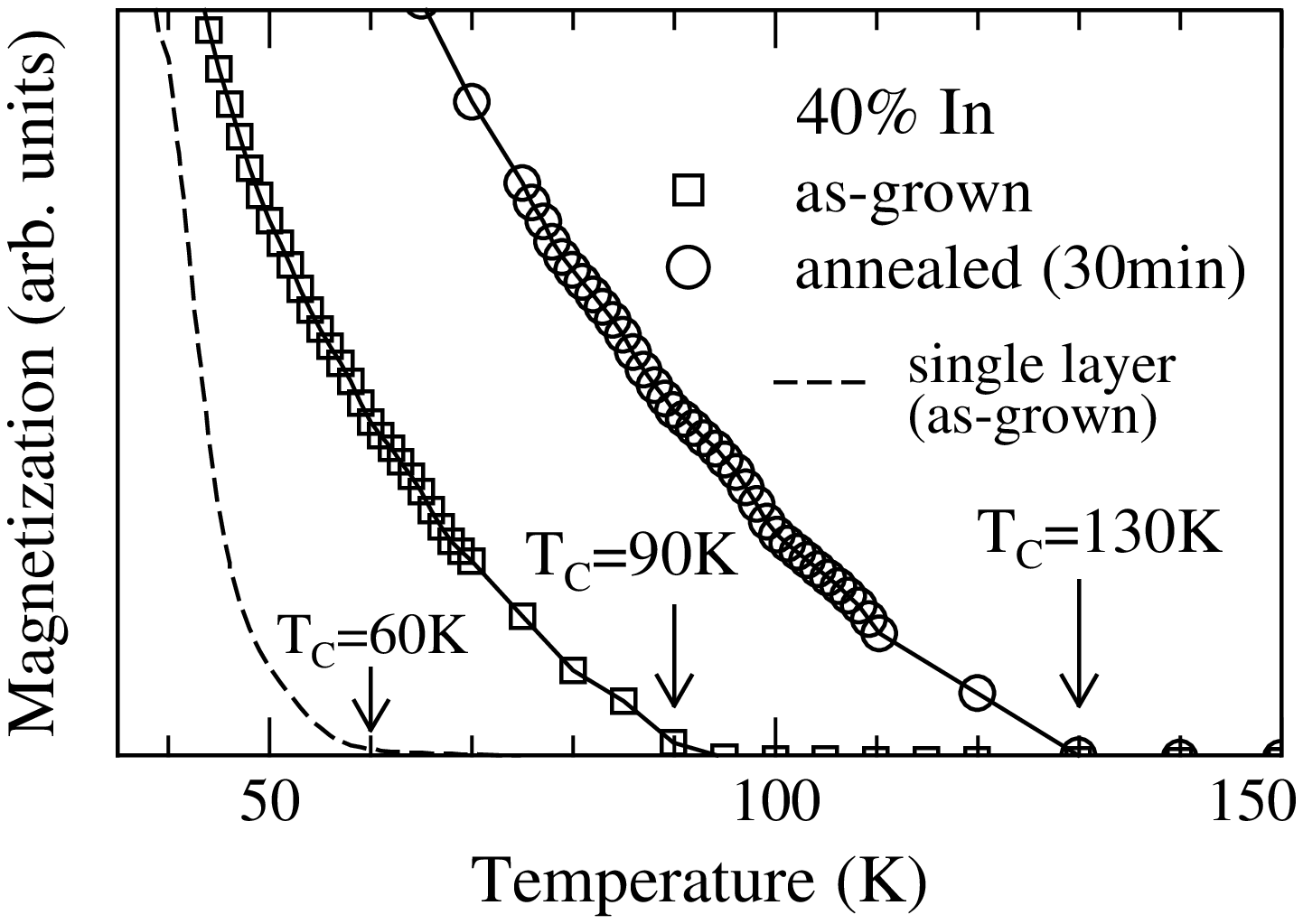}
{\bf Fig.~6}: Magnetization of a GaMnAs/InGaMnAs heterostructure as a function of temperature
before and after annealing.
\end{center}
Curie temperatures of 90 K and 130 K are determined from the SQUID measurements, respectively.
Thus, an enhancement of 30 K has been achieved in comparison to the GaMnAs single layer. Similar to the case of the
GaMnAs layer, electrical in-situ measurements during annealing show a monotonous increase of the conductivity with
increasing annealing time, whereas the Raman spectra and the SQUID measurements reveal that no further increase of the
hole density and the Curie temperature, respectively, is achieved by annealing for more than 30 minutes.
The enhancement of the Curie temperature strongly correlates with the In concentration. More details will be
discussed elsewhere.

\vspace{0.3cm}
{\bf 4. Conclusions}

\vspace{0.2cm}
\hspace*{0.3cm}
We have shown that annealing of GaMnAs single layers with thicknesses larger than 300 nm
and GaMnAs/InGaMnAs heterostructures at 250$^\circ$C
for more than 6 h leads to a monotonous increase of the conductivity with increasing annealing time. Consequently,
Raman and SQUID measurements reveal an enhancement of the hole density and $T_C$, respectively.
However, this enhancement drastically exceeds that expected from the conductivity data. Furthermore, a saturation of the
hole density and $T_C$ is observed after annealing for 30 minutes. The discrepancies in the results,
obtained from the conductivity measurements on the one hand and those obtained from the Raman and SQUID measurements
on the other hand, are attributed to a vertical gradient of the hole density and the mobility within the GaMnAs
layers. The incorporation of thin InGaMnAs films results in a drastic increase of the Curie temperature of the
GaMnAs layers.

\vspace{0.2cm}

{\bf References}

[1] T. Dietl, H. Ohno, and F. Matsukura, Phys. Rev. B 63 (2001) 195205.\newline
[2] K.C. Ku, S.J. Potashnik, R.F. Wang, S.H.Chun, P. Schiffer, N. Samarth, M.J. Seong, A. Mascarenhas,
E. Johnston-Halperin, R.C. Myers, A.C. Gossard, and D.D. Awschalom, Appl. Phys. Lett. 82 (2003) 2302.\newline
[3] S.J. Potashnik, K.C. Ku, S.H.Chun, J.J. Berry, N. Samarth, and P. Schiffer, Appl. Phys. Lett. 79 (2001)
1495.\newline
[4] K.W. Edmonds, K.Y. Wang, R.P. Campion, A.C. Neumann, N.R.S. Farley, B.L. Gallagher, and C.T. Foxon,
Appl. Phys. Lett. 81 (2002) 4991.\newline
[5] B.S. S{\o}rensen, P.E. Lindelof, J. Sadowski, R. Mathieu, and P. Svedlindh, Appl. Phys. Lett. 82 (2003) 2287.
\newline
[6] D. Ruzmetov, J. Scherschligt, D.V. Baxter, T. Wojtowicz, X. Liu, Y. Sasaki, J.K. Furdyna, K.M. Yu,
and W. Walukiewicz, cond-mat/0302013.\newline
[7] W. Limmer, M. Glunk, S. Mascheck, A. Koeder, D. Klarer, W. Schoch, K. Thonke, R. Sauer, and A. Waag,
Phys. Rev. B 66 (2002) 205209.\newline
[8] A. Koeder, S. Frank, W. Schoch, V. Avrutin, W. Limmer, K. Thonke, R. Sauer, A. Waag, M. Krieger, K. Zuern,
P. Ziemann, S. Brotzmann, and H. Bracht, Appl. Phys. Lett. 82 (2003) 3278. \newline
[9] J. Blinowski and P. Kacman, Phys. Rev. B 67 (2003) 121204.\newline
\end{document}